\documentclass[reprint,aps,prl]{revtex4-1}

\usepackage{graphicx}
\usepackage{natbib}
\usepackage{hyperref}

\newcommand{\bra}[1]{\ensuremath{\langle #1 |}}
\newcommand{\ket}[1]{\ensuremath{| #1 \rangle}}

\newcommand{\ii}{\ensuremath{\mathrm{i}}}

\begin{document}

\title{Experimental High-Dimensional Entanglement by Path Identity}

\author{Jaroslav Kysela}
\email{jaroslav.kysela@univie.ac.at}
\affiliation{Institute for Quantum Optics and Quantum Information (IQOQI), Austrian Academy of Sciences, Boltzmanngasse 3, A-1090 Vienna, Austria}
\affiliation{Faculty of Physics, University of Vienna, Boltzmanngasse 5, 1090 Vienna, Austria}

\author{Manuel Erhard}
\email{manuel.erhard@univie.ac.at}
\affiliation{Institute for Quantum Optics and Quantum Information (IQOQI), Austrian Academy of Sciences, Boltzmanngasse 3, A-1090 Vienna, Austria}
\affiliation{Faculty of Physics, University of Vienna, Boltzmanngasse 5, 1090 Vienna, Austria}

\author{Armin Hochrainer}
\affiliation{Institute for Quantum Optics and Quantum Information (IQOQI), Austrian Academy of Sciences, Boltzmanngasse 3, A-1090 Vienna, Austria}
\affiliation{Faculty of Physics, University of Vienna, Boltzmanngasse 5, 1090 Vienna, Austria}

\author{Mario Krenn}
\altaffiliation[Present address: ]{Department of Chemistry, University of Toronto, Toronto, ON M5S 3H6, Canada}
\affiliation{Institute for Quantum Optics and Quantum Information (IQOQI), Austrian Academy of Sciences, Boltzmanngasse 3, A-1090 Vienna, Austria}
\affiliation{Faculty of Physics, University of Vienna, Boltzmanngasse 5, 1090 Vienna, Austria}

\author{Anton Zeilinger}
\email{anton.zeilinger@univie.ac.at}
\affiliation{Institute for Quantum Optics and Quantum Information (IQOQI), Austrian Academy of Sciences, Boltzmanngasse 3, A-1090 Vienna, Austria}
\affiliation{Faculty of Physics, University of Vienna, Boltzmanngasse 5, 1090 Vienna, Austria}

\date{\today}

\begin{abstract}
Versatile and high-brightness sources of high-dimensional entangled photon pairs are important for emerging quantum technologies such as secure quantum communication. Here, we experimentally demonstrate a new scalable method to create photon pairs carrying orbital angular momentum that are entangled in arbitrarily high dimensions. Our method relies on indistinguishable photon pairs created coherently in different sources. We demonstrate the creation of three-dimensionally entangled states and show how to incrementally increase the dimensionality of entanglement. The generated states retain their quality even in higher dimensions. In addition, the modular structure of our approach allows for generalization to various degrees of freedom and even implementation in integrated compact devices. We therefore expect that future quantum technologies and fundamental tests of nature in higher dimensions will benefit from this novel approach.
\end{abstract}

\maketitle

\emph{Introduction.}---The transition from two- to multi-dimensional entangled quantum systems brings about radical improvements in the distribution and processing of quantum information. Such systems play an important role in secure communication over high capacity quantum channels \cite{superdensecoding}. They offer improved noise resistance and increased security against eavesdropping \cite{weakrandom1,cerf2002security}. Furthermore, multi-dimensionally entangled quantum systems are beneficial for experiments that are interesting from a more fundamental perspective, such as tests of local realism \cite{cglmp,ghznqudits,ghzrotcov,detectionloophole2}. If we think of teleporting the entire information of a single or even several photons in the future \cite{multiteleportation,malik2016multi,erhard2018experimental}, such multi-dimensionally entangled sources will be indispensable. In view of these perspectives and applications, high quality sources for multi-dimensional entanglement are highly desirable.

Various degrees of freedom, such as frequency \cite{onchipent}, time-bin \cite{thew,deRiedmatten,tiranov} and path \cite{christophchip,pathhighdim}, have been employed so far for the generation of high-dimensionally entangled states. In the following, we focus on the orbital angular momentum (OAM) of photons, but our technique is valid for other degrees of freedom as well. The OAM of photons is an in principle unbounded discrete quantity and as such has been used extensively \cite{oamorig,molinareview,roadmapoam,padgettoverview,manuelreview} to prepare high-dimensionally entangled photonic states. In the traditional way, the OAM-entangled photon pairs are produced in a single spontaneous parametric down-conversion (SPDC) process \cite{mair}. Albeit convenient, this process exhibits several drawbacks. For example, photon pairs generated in this way have a non-uniform distribution of OAM \cite{spiralbandwidththeory,dada,spiralbandwidthexper,marioent100}. The maximally entangled states can then be generated either by post-processing techniques, such as Procrustean filtering \cite{procrusteanfilt,vaziriProcrustean}, or by pre-processing of the pump beam. In a recently demonstrated approach \cite{kovlakov,liu}, a superposition of OAM modes is imprinted by holograms into the pump beam, which translates via down-conversion into maximally entangled states of two photons.

In this paper we present the first ex\-per\-i\-men\-tal dem\-on\-stra\-tion of a new method of generating high-di\-men\-sion\-al\-ly entangled states. Multiple SPDC processes are employed, but none of them directly produces entangled states. The method relies on the concept known as \emph{entanglement by path identity} \cite{marioepi,mariograph}, which was discovered recently with the help of a computer program \cite{melvin}. Our technique displays several important advantages over the traditional approach. Our source of entangled photon pairs is versatile as both magnitudes and phases in a high-dimensional quantum state can be adjusted completely arbitrarily. Furthermore, the experimental implementation of our source has a modular structure. The modularity, together with the fact that adding one dimension to the entangled state increases the count rates accordingly, makes our approach scalable. High brightness of our source is ensured as all photons are produced already in the desired modes and no photons have to be discarded by post-selection. 

This work is organized as follows. After a brief introduction to the concept of entanglement by path identity, we describe the main features of our experimental implementation. Then we demonstrate the scalability and versatility of our method by generating several different states in two and three dimensions. We verify the quality of the produced entangled quantum states using quantum state tomography.

\begin{figure}
\centering
\includegraphics[scale=.32]{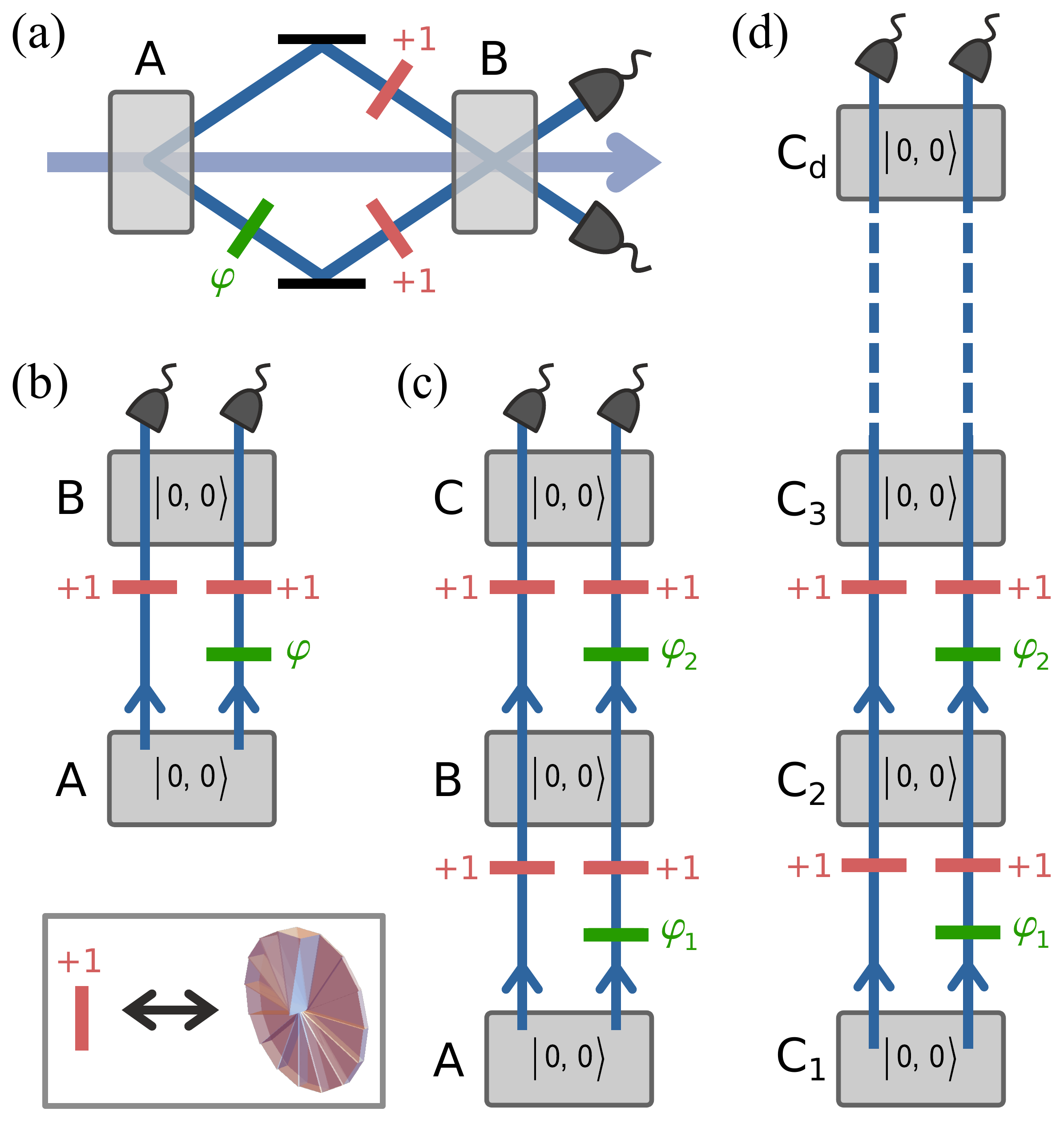}
\caption{\label{fig:principle}(color online). Basic concept. Gray boxes labeled with capital letters represent nonlinear crystals, each pumped coherently and each generating with a small probability a pair of photons via an SPDC process. Each generated photon-pair is in state $\ket{0,0}$ with a small contribution of higher-order terms. The two down-converted photons then propagate along their paths in the direction indicated by the arrows and acquire phase shifts $\varphi_i$ as well as additional quanta of OAM due to phase- and mode-shifters. (a) The pump beam, represented by an arrow, gives rise to an SPDC process in crystals A and B. Photons generated in crystal A are reflected into crystal B such that their paths are overlapped with paths of photons generated in crystal B. As a consequence, the two coherent SPDC processes in crystals A and B are indistinguishable and the generated photon pairs leave the setup in a two-dimensionally entangled Bell state $1/\sqrt{2} \, (\ket{0,0} + \exp{(i \varphi)} \ket{1,1})$. The quantum of OAM is imparted to the photon by a spiral phase plate shown in the inset. (b) The schematic picture of the setup in (a), where the pump beam is not shown. (c) The addition of the third crystal to the setup increases the entanglement dimension by one. The resulting state thus reads $1/\sqrt{3} \, (\ket{0,0} + \exp{(i \bar{\varphi}_1)} \ket{1,1} + \exp{(i \bar{\varphi}_2)} \ket{2,2})$, where $\bar{\varphi}_1 = \varphi_2$ and $\bar{\varphi}_2 = \varphi_1 + \varphi_2$. (d) One can stack multiple setups from (a) to acquire a series of $d$ crystals that produces a $d$-dimensionally entangled state $1/\sqrt{d}(\ket{0,0} + \exp{(i \bar{\varphi}_1)} \ket{1,1} + \ldots + \exp{(i \bar{\varphi}_{d-1})} \ket{d-1,d-1})$, where the relative phases $\bar{\varphi}_i = \sum_{j=d-i}^{d-1} \varphi_j$ are adjusted by an appropriate choice of phase-shifters $\varphi_j$. The magnitudes of the individual modes are modified by varying the power with which the respective crystals are pumped.
}
\end{figure}

\label{sec:epi}
\emph{Entanglement by path identity.}---Consider a simple experimental setup consisting of two nonlinear crystals that are aligned in series and coherently emit photons via SPDC, as shown in Fig. \ref{fig:principle}(a) and (b). The pump power for both crystals is set sufficiently low such that events when either crystal emits multiple photon-pairs as well as events when both crystals simultaneously generate a photon-pair can be neglected. The propagation paths of the down-converted photons coming from the two crystals are carefully overlapped. As a result, once the photon pair leaves the setup, no information can be obtained, not even in principle, in which crystal the pair was created \cite{wang,herzog,herzog1995}. The down-conversion processes in both crystals are adjusted such that photon pairs may be emitted only into the fundamental mode $\ket{0,0}$ with zero quanta of OAM \footnote{Parameters of the two SPDC processes are chosen such that photon properties such as frequency, polarization, and OAM are identical for both crystals and also higher-order OAM modes are highly suppressed.}. Importantly, no entanglement is generated by either of the two crystals. (In practice, a small contribution of higher-order OAM modes is also present. For the detailed discussion see Supplementary \cite{supp}.)

Suppose now that two mode-shifters are inserted into the setup. These add an extra quantum of OAM to each photon originating in the first crystal and thus act as the only source of which-crystal information. As the down-conversion processes in the two crystals are (apart from the OAM) indistinguishable, the resulting state of a detected photon pair is a coherent superposition
\begin{equation}
    \ket{\psi} = \frac{1}{\sqrt{2}}(\ket{0,0} + e^{i \varphi} \ket{1,1}).
\end{equation}
In the formula above, $\varphi$ is the phase between the two SPDC processes imparted by a phase-shifter and numbers in ket vectors refer to the OAM quanta of respective photons.

The generation of entangled states as described above is a specific example of the concept termed entanglement by path identity. This concept can be readily generalized for production of high-dimensionally entangled states \cite{marioepi}. When the number of crystals in the series is increased to $d$, and the number of phase- and mode-shifters is accordingly increased to $d - 1$, high-dimensionally entangled states of the following form are produced
\begin{equation}
\ket{\psi} = \sum_{\ell=0}^{d-1} c_\ell \ket{\ell,\ell},
\label{eq:epi_state}
\end{equation}
where $d$ is the state dimension and $c_\ell$ are complex amplitudes, see Fig.~\ref{fig:principle}(c) and (d). The magnitudes of $c_\ell$ can be set by pumping each crystal independently with properly adjusted power. By using different mode-shifters for either of the two photons in a down-converted pair, completely arbitrary states can be created. Interestingly, the widely used cross-crystal scheme is the simplest example of the above approach, where two-particle states are entangled in polarization \cite{hardy,kwiatsource,twelvephotons}.

\begin{figure*}
\includegraphics[scale=.4]{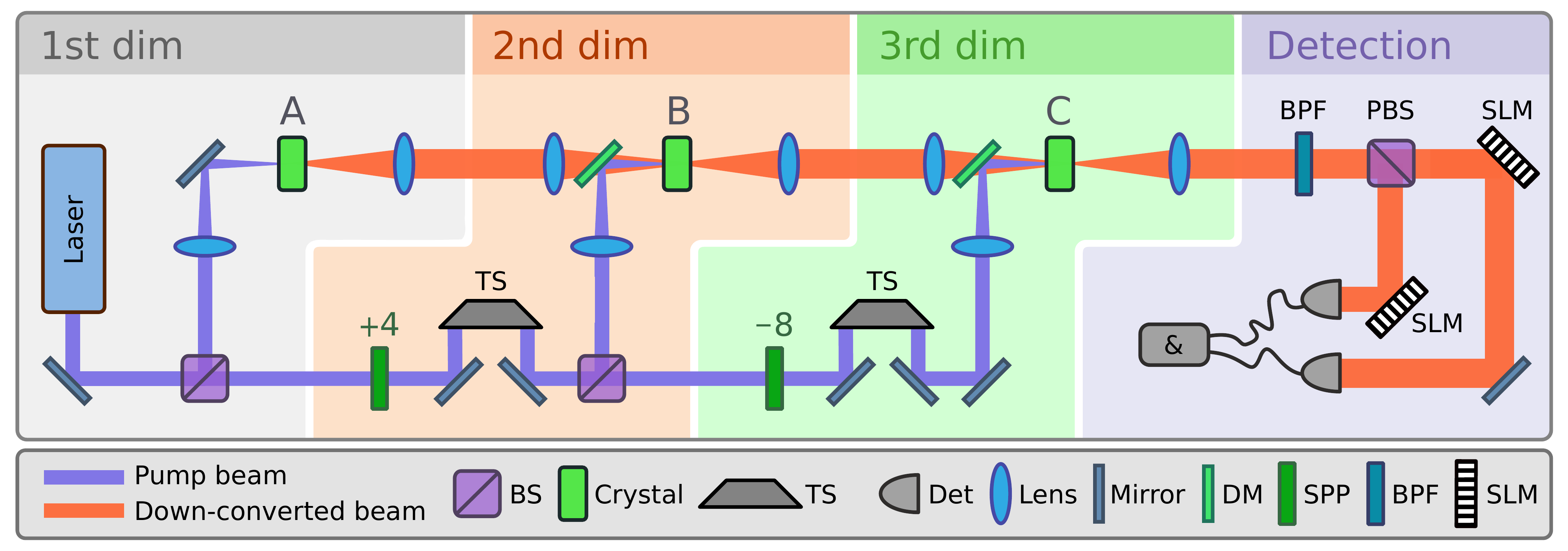}
\caption{\label{fig:setup}(color online). Experimental setup. Three-dimensional states are created by elements in boxes \textsf{1st dim}, \textsf{2nd dim} and \textsf{3rd dim}. Three periodically poled KTP crystals A, B and C are pumped with a continuous-wave laser beam at the central wavelength of 405~nm. Frequency-degenerate down-converted photons created by type II collinear SPDC propagate along identical paths into the detection system shown in box \textsf{Detection}. Photons originating in crystal B are created in $\ket{2, 2}$ OAM mode because of a spiral phase plate (SPP $+4$) inserted after the first beam-splitter (BS). In addition, photons originating in crystal C are created in $\ket{-2,-2}$ mode due to an extra mirror that effectively works as a $-8$ mode-shifter as is explained in the main text. The pump beam is separated from the down-conversion beam by dichroic mirrors (DM) and a band-pass filter (BPF). Before detection, the two down-converted photons are separated on a polarizing beam-splitter (PBS). The state tomography in the OAM degree of freedom is done by projective measurements \cite{mair} where specific holograms are projected on two spatial light modulators (SLMs). The reflected photons are subsequently coupled into single mode fibers and detected by single photon counting modules (Det). The resulting signals are post-processed by a coincidence counting module (\&). The relative phases $\varphi_1$ and $\varphi_2$ can be adjusted by phase-shifters implemented with trombone systems (TS). The magnitudes of individual terms in the quantum state are controlled by setting the splitting ratio of the beam-splitters. For the detailed diagram of the experimental setup see Supplementary \cite{supp}.}
\end{figure*}

\label{sec:setup}
\emph{Setup.}---The experimental implementation presented here is based on the scheme in Fig.~\ref{fig:principle}(c) with two main modifications. First, the pump and down-converted beams for each crystal are separated by two Mach-Zehnder interferometers. This way, phases as well as magnitudes of individual modes in the quantum state can be adjusted independently. In addition, there is an advantage that both wavelengths can be manipulated separately. Second, the mode-shifter is placed into the pump beam instead of the down-conversion beam. This measure was taken to preclude the mode-shifter from introducing undesirable OAM terms into the quantum state. For technical reasons, the down-converted photon pairs were not emitted in a perfectly collinear manner, but had a slight angular spread of roughly $1^\circ$. This leads to a non-perfect operation of the mode-shifter, which functions properly only when both photons propagate through its center (for details see~\cite{[{See Supplemental Material for the detailed description of the experimental setup, spiral spectra of nonlinear crystals and complete state tomography data}]supp}).

The setup, presented in Fig.~\ref{fig:setup}, was designed to produce three-dimensionally entangled states. Each dimension in the generated quantum state corresponds to one of three nonlinear crystals A, B or C in the setup. In Fig.~\ref{fig:setup} this correspondence is emphasized by enclosing the crystals with associated elements into boxes \textsf{1st dim}, \textsf{2nd dim} and \textsf{3rd dim}. The laser beam is split into three paths to pump each crystal separately. The pump beam for crystal A possesses zero quanta of OAM and so do the down-converted photons \cite{supp}, which exit the crystal in state $\ket{0,0}$. (Apart from the predominant $\ket{0, 0}$ component, also small but negligible contributions of higher-order OAM correlations are present in the photons' state. For details see Supplementary \cite{supp}.) The pump beam for crystal B acquires four quanta of OAM due to a spiral phase plate (SPP), which is inserted into the beam and plays the role of the mode-shifter. Consequently, each down-converted photon generated in crystal B carries two quanta of OAM and the pair is produced in state $\ket{2,2}$. Similarly, the pump beam for crystal C also acquires four quanta of OAM, but an additional mirror is used to invert the sign of the OAM value from $4$ to $-4$, effectively subtracting eight quanta of OAM. Down-converted photons coming from crystal C are then produced in state $\ket{-2,-2}$. The resulting quantum state reads
\begin{equation}
\ket{\psi} = \alpha \! \! \underbrace{\ket{0,0}}_{\text{crystal A}} + \ \beta e^{i\varphi_1} \! \! \! \underbrace{\ket{2,2}}_{\text{crystal B}} + \ \gamma e^{i\varphi_2} \underbrace{\ket{-2,-2}}_{\text{crystal C}}.
\label{eq:3dstate}
\end{equation}
Magnitudes $\alpha$, $\beta$ and $\gamma$ of the entangled state can be changed by adjusting the relative pump power for each crystal. The relative phases $\varphi_1$ and $\varphi_2$ are set by positioning two trombone systems that act as phase-shifters. By employing only the first two stages of the setup, namely parts in boxes \textsf{1st dim} and \textsf{2nd dim}, two-dimensionally entangled states are created.

We use type II SPDC in all three crystals. In order to measure the entangled state, we first deterministically separate the two down-converted photons by a polarizing beam splitter. Two spatial light modulators in combination with single mode fibers are used to perform any projective measurement for OAM modes \cite{mair}. The single photons are then detected by avalanche photon detectors and simultaneous two-photon events are identified by a coincidence logic.

Finally, the resulting quantum states are characterized by complete quantum state tomography. We use a maximum-likelihood reconstruction technique \cite{hradilMLE} to estimate the physical density matrices of the detected photon pairs.

\begin{figure*}[htbp]
	\centering
	\includegraphics[width=0.99\textwidth]{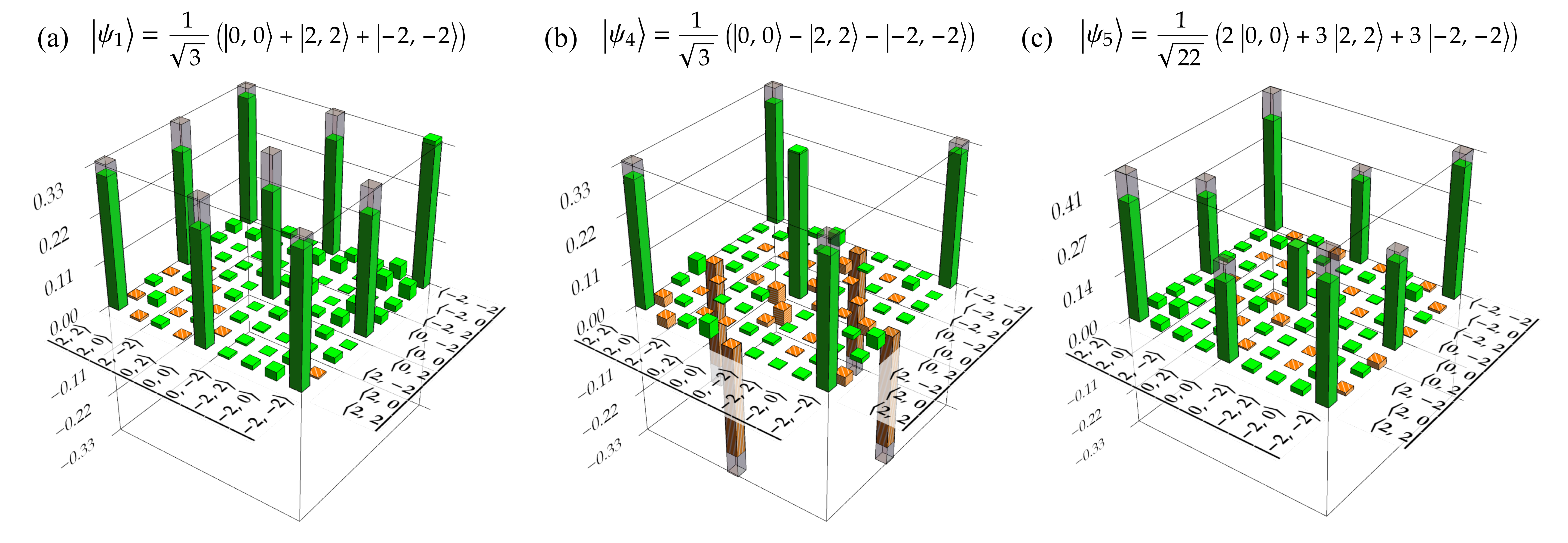}
	\caption{(color online). Examples of produced three-dimensionally entangled states $\ket{\psi_1}$, $\ket{\psi_4}$ and $\ket{\psi_5}$, see Table~\ref{tab:fidelities}. With our method we can control the relative phases, as demonstrated in (a) and (b), as well as relative magnitudes, as can be seen in (c). Only real parts are shown, imaginary parts lie in the range $(-0.12, 0.12)$ for all cases. Green (solid) and orange (hatched) bars represent positive and negative values of reconstructed density matrices, respectively. Gray translucent bars represent the theoretical expectation. Fidelities of the measured states with their reference states are $87.0 \pm 0.5 \%$, $89.0 \pm 0.4 \%$ and $84.8 \pm 0.8 \%$, respectively.}
	\label{fig:3d-density-matrices}
\end{figure*}

\label{sec:results}
\emph{Experimental Results.}---The high flexibility of our setup in producing various states is demonstrated in Tab.~\ref{tab:fidelities}, where fidelities for different three-dimensionally (and also two-dimensionally) entangled states are presented. These data demonstrate our ability to control the relative phases and magnitudes of the generated quantum states. Most notably, we are able to create three mutually orthogonal and maximally entangled states $\ket{\psi_1}$, $\ket{\psi_2}$ and $\ket{\psi_3}$ with an average fidelity of $87.5 \pm 0.6 \%$. The orthogonality of these states does not follow directly from the orthogonality of OAM modes, but indeed from differently adjusted phases in the quantum states. With the state $\ket{\psi_5}$ we demonstrate the ability to adjust relative magnitudes of terms in the quantum superposition. We do so by slightly reducing the probability amplitude for the zero OAM mode. The real parts of density matrices for three of the states presented in Tab.~\ref{tab:fidelities} are displayed in Fig.~\ref{fig:3d-density-matrices}. There the measurement results (solid bars) are compared to the theoretical expectations (translucent bars). The average fidelity of three-dimensionally entangled states does not decrease significantly when compared to the average fidelity of $89.8 \pm 0.5 \%$ for two-dimensional states \cite{supp}. The quality of entangled states is thus mostly unaffected when going from two to three dimensions and indicates the scalability of our approach for even higher dimensions. The main limitations for achieving higher coherence are slight distinguishability of the SPDC sources, which we attribute to small misalignments and imperfect transformations of OAM modes. However, none of these imperfections are of fundamental nature and we discuss possible improvements of our design in the following section.

\begin{table}
\caption{\label{tab:fidelities} Fidelities $F(\ket{\psi},\rho)=\text{Tr}(\ket{\psi}\bra{\psi}\rho)$ between several two- and three-dimensionally entangled states $\ket{\psi}$ and their experimental realizations $\rho$. States $\ket{\psi_1}$, $\ket{\psi_2}$ and $\ket{\psi_3}$ form an orthonormal set of maximally entangled states in three dimensions ($\omega = e^{2 \pi \ii/3}$). State $\ket{\psi_5}$ is a manifestation of our ability to control not only relative phases in the quantum state, but also relative magnitudes.}
\begin{center}
   \begin{tabular}{| l || c |}
     \hline
       State & Fidelity $F$ \\ \hline \hline
       
       $\ket{\Phi^+} = 1/\sqrt{2}(\ket{0,0}+\ket{2,2})$ & $0.904 \pm 0.005$ \\ \hline
		$\ket{\Phi^-} = 1/\sqrt{2}(\ket{0,0}-\ket{2,2})$ & $0.891 \pm 0.005$ \\ \hline \hline

       $\ket{\psi_1}=\frac{1}{\sqrt{3}}(\ket{0,0}+\ket{2,2}+\ket{-2,-2})$ & $0.870\pm0.005$ \\ \hline
       $\ket{\psi_2}=\frac{1}{\sqrt{3}}(\ket{0,0}+\omega \ket{2,2}+\omega^{-1} \ket{-2,-2})$ & $0.852\pm0.007$ \\ \hline
       $\ket{\psi_3}=\frac{1}{\sqrt{3}}(\ket{0,0}+\omega^{-1} \ket{2,2}+ \omega \ket{-2,-2})$ & $0.903\pm0.006$ \\ \hline
       $\ket{\psi_4}=\frac{1}{\sqrt{3}}(\ket{0,0}-\ket{2,2}-\ket{-2,-2})$ & $0.890\pm0.004$ \\ \hline
       $\ket{\psi_5}=\frac{1}{\sqrt{22}}(2 \ket{0,0}+3 \ket{2,2}+3 \ket{-2,-2})$ & $0.848\pm0.008$ \\ \hline
   \end{tabular}
\end{center}
\end{table}

\label{sec:scalability}
\emph{Scalability.}---The scalability of our scheme is enabled by the modular structure of the experimental setup. Adding a crystal and a mode-shifter results in an increase of the entanglement dimension by one. In order to further improve the scalability, some modifications to our experimental implementation can be made. We adopted the Mach-Zehnder interferometric configuration in our experiment. This gives us freedom to access and manipulate the pump and down-conversion beams separately with no need of custom-made components. Currently, the size of the setup is governed by the two 4f lens systems between the crystals. The distance between two successive crystals in our setup is 600~mm. Due to these large interferometers active stabilization is inevitable. However, scaling down the distances and employing integrated fabrication techniques as used in micro-chip fabrication leads to intrinsically stable interferometers. An alternative approach is to circumvent interferometers completely. For example, by utilizing wavelength-dependent phase-shifters and q-plates \cite{qplate}. This approach leaves the pump beam unaffected while manipulating only the down-conversion beam \cite{wavedepqplate}. Thus the pump beam can co-propagate along the same path as the down-conversion photons leading to intrinsic phase stability.

\label{sec:summary}
\emph{Conclusion.}---In conclusion, we experimentally demonstrated entanglement by path identity \cite{marioepi} as a new method of generating high-dimensionally entangled states. The OAM degree of freedom of photons was utilized in our setup. We showed the flexibility of our approach by producing various entangled quantum states in two and in three dimensions. Both magnitudes and phases of complex amplitudes in the quantum states can be adjusted to arbitrary values independently of each other. The experimental setup has a modular structure and in principle allows to generate entangled states in arbitrarily high dimensions.

The modularity of the setup and the fact that only three kinds of elements---crystals, mode- and phase-shifters---are needed is the core strength of our approach. It promises high brightness, scalability in the dimension, and versatility of generated states. A very appealing feature of our method is that different families of spatial modes can be utilized. It is therefore possible to generate high-dimensionally entangled photon pairs in specialized modes, which are optimized, for example, for free-space communication or even for fiber-based systems.\nocite{coherencecond1,coherencecond2,miatto}
\\
\\
This work was supported by the Austrian Academy of
Sciences (OeAW), the European Research Council (SIQS
Grant No. 600645 EU-FP7-ICT), and the Austrian Science
Fund (FWF) with SFB F40 (FoQuS) and FWF project W 1210-N25 (CoQuS).
\\
\\
J. K. and M. E. contributed equally to this work.

\bibliography{references}

\end{document}


\title{Experimental High-Dimensional Entanglement by Path Identity: Supplemental Material}

\author{Jaroslav Kysela}
\email{jaroslav.kysela@univie.ac.at}
\affiliation{Institute for Quantum Optics and Quantum Information (IQOQI), Austrian Academy of Sciences, Boltzmanngasse 3, A-1090 Vienna, Austria}
\affiliation{Faculty of Physics, University of Vienna, Boltzmanngasse 5, 1090 Vienna, Austria}

\author{Manuel Erhard}
\email{manuel.erhard@univie.ac.at}
\affiliation{Institute for Quantum Optics and Quantum Information (IQOQI), Austrian Academy of Sciences, Boltzmanngasse 3, A-1090 Vienna, Austria}
\affiliation{Faculty of Physics, University of Vienna, Boltzmanngasse 5, 1090 Vienna, Austria}

\author{Armin Hochrainer}
\affiliation{Institute for Quantum Optics and Quantum Information (IQOQI), Austrian Academy of Sciences, Boltzmanngasse 3, A-1090 Vienna, Austria}
\affiliation{Faculty of Physics, University of Vienna, Boltzmanngasse 5, 1090 Vienna, Austria}

\author{Mario Krenn}
\altaffiliation[Present address: ]{Department of Chemistry, University of Toronto, Toronto, ON M5S 3H6, Canada}
\affiliation{Institute for Quantum Optics and Quantum Information (IQOQI), Austrian Academy of Sciences, Boltzmanngasse 3, A-1090 Vienna, Austria}
\affiliation{Faculty of Physics, University of Vienna, Boltzmanngasse 5, 1090 Vienna, Austria}

\author{Anton Zeilinger}
\email{anton.zeilinger@univie.ac.at}
\affiliation{Institute for Quantum Optics and Quantum Information (IQOQI), Austrian Academy of Sciences, Boltzmanngasse 3, A-1090 Vienna, Austria}
\affiliation{Faculty of Physics, University of Vienna, Boltzmanngasse 5, 1090 Vienna, Austria}

\maketitle

\section{Detailed setup}

The simplified scheme of the experimental setup shown in Fig.~2 in the main text does not show the implementation of the variable splitting-ratio beam-splitters, phase shifters, and the interferometer stabilization system. In Fig.~\ref{fig:detailed_setup} the detailed scheme of our setup can be found. The variable splitting-ratio beam-splitter is built from a polarizing beam-splitter and two half-wave plates. After each beam-splitter both beams possess horizontal polarization.

\begin{figure}[htbp]
	\includegraphics[width=0.45\textwidth]{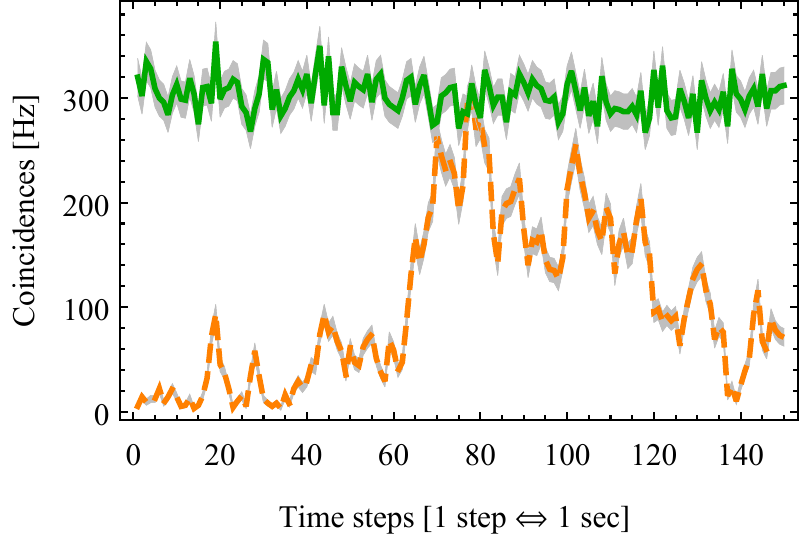}
	\caption{\label{fig:phase_stability}Comparison of coincidence-count signal fluctuations for the case when the interferometer in the setup is actively stabilized (green solid line) and when it is not (orange dashed line). Photon pairs coming from crystals A and B, pumped with a beam having zero quanta of OAM, were collected in time steps of one second. Gray shaded areas correspond to one standard deviation region of collected data when Poissonian counting statistics is assumed.}
\end{figure}

To phase-stabilize the interferometer in the setup, a light beam emitted from an additional laser at the central wavelength of 710~nm is injected into the unused input port of the beam-splitter. The two created spatial modes of light are recombined at the dichroic mirror and the resulting interfering beam is monitored by a fast photodiode. The detection signal is processed by a PID controller and a feedback signal is fed to the piezo-actuator-driven mirror in order to compensate for phase fluctuations. For comparison, in Fig.~\ref{fig:phase_stability} the time dependence of the coincidence count rate with and without stabilization is presented. This active feedback loop stabilization system also works as a fine phase shifter, for details see section ``Phase adjustment.'' The coarse adjustment of the phase can be done by a trombone system built in one arm of the interferometer.

The OAM mode shifter is inserted into the pump beam instead of the down-converted beams, as suggested by the principle scheme in Fig.~1 in the main text, for the following reason. Requirements on the temperature of nonlinear crystals and on the frequencies of down-converted photons did not allow for perfect collinearity of the generated pairs. In order to operate optimally the mode shifter has to be precisely centered with respect to the beam it acts upon. The failure to satisfy this condition for both photons of the pair leads to generation of undesirable higher-order OAM terms and a spread of the resulting OAM spiral spectrum. 

\begin{figure}
\includegraphics[scale=1]{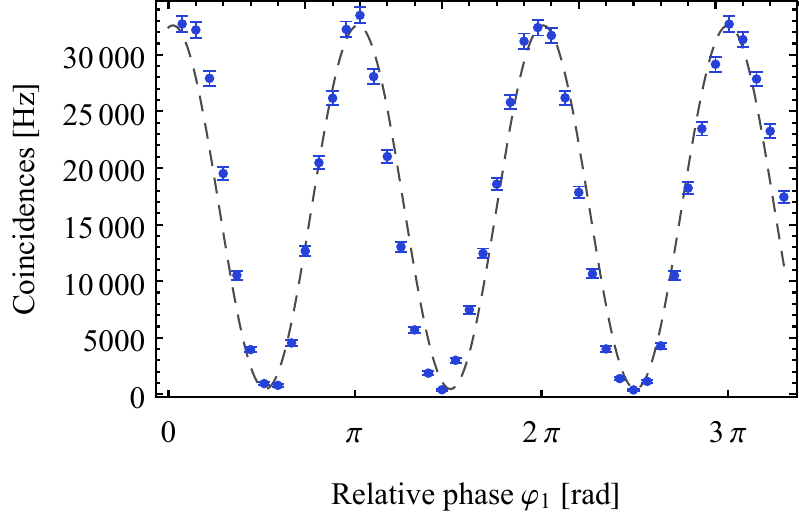}
\caption{\label{fig:interference}Interference of two SPDC processes in crystals A and B in the fundamental OAM mode $\ket{0,0}$. Coincidence counts are collected while the relative phase $\varphi_1$ is changed. Errors are determined assuming Poissonian counting statistics. The dashed line represents the fit of the displayed data. The obtained visibility is $97.1 \pm 0.5 \%$.}
\end{figure}

\begin{figure*}[htbp]
	\includegraphics[width=1\textwidth]{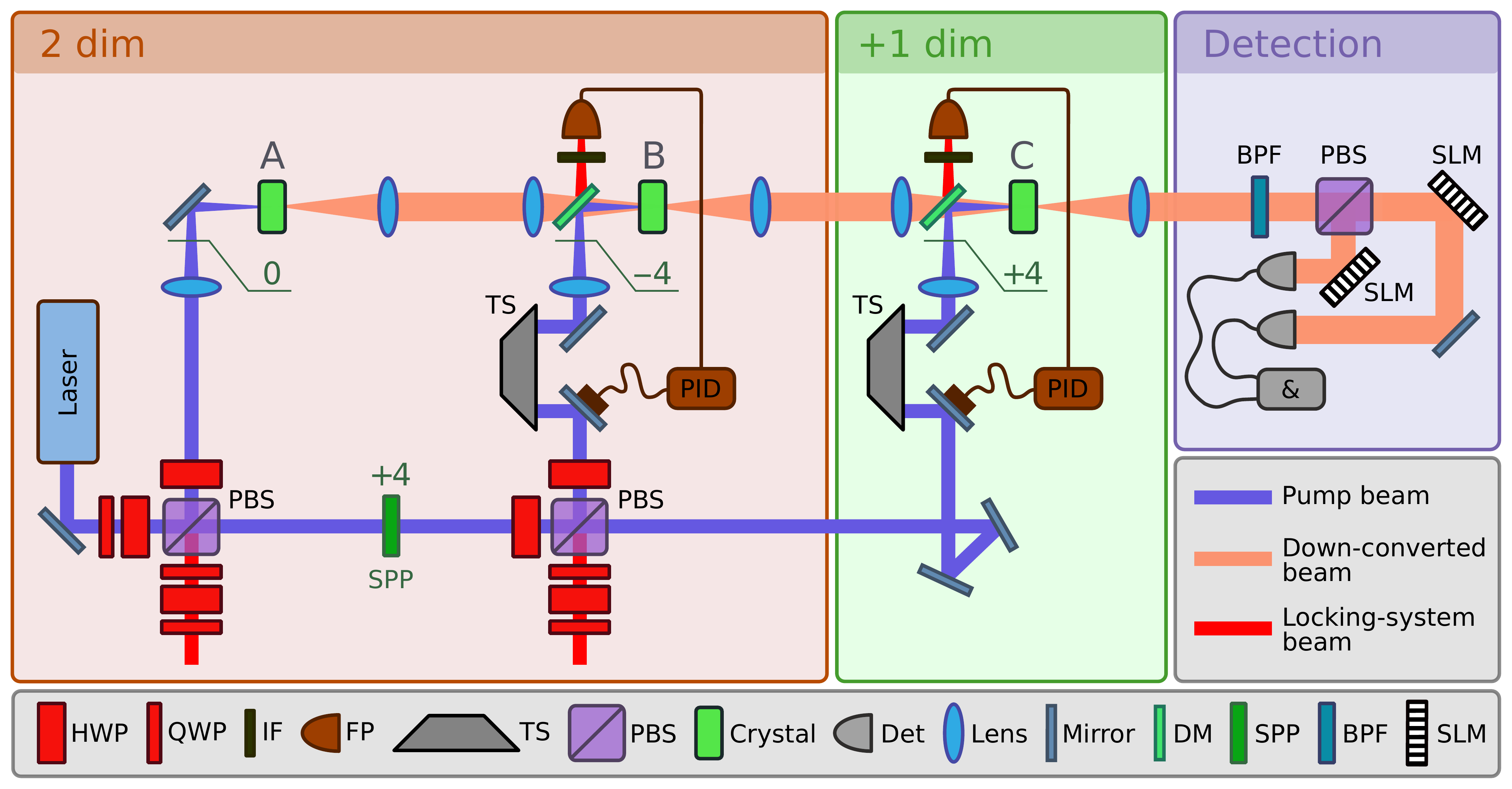}
	\caption{\label{fig:detailed_setup}Detailed setup. For a description of the basic setup see Fig.~2 in the main text. The two interferometers are phase-locked by active feedback systems. An additional laser diode with central wavelength 710~nm is used to provide a locking-system beam that is injected into unused ports of beam-splitters and leaves the setup through dichroic mirrors (DM). After filtering out the pump beam by an interference filter (IF), the interference fluctuations of the locking-laser beam are detected by a fast photodiode (FP). The obtained signal is processed by a PID controller and a feedback signal is sent to a piezo actuator attached to one of the mirrors in the interferometer. The relative phases of the down-converted beams (denoted by $\varphi_1$ and $\varphi_2$ in the main text) can be adjusted by trombone systems (TS) and by a proper setting of polarization of the corresponding locking-system beam. This is accomplished by a series of two quarter-wave plates (QWP) and one half-wave plate (HWP) as is explained in the text. Magnitudes of individual terms in the quantum state are controlled by setting the splitting ratio of the beam-splitters. A variable splitting-ratio beam-splitter is implemented by a polarizing beam-splitter (PBS) with two half-wave plates. All three crystals are 10 mm long ppKTP non-linear crystals. Det -- single photon detector, SPP -- spiral phase plate, BPF -- band-pass filter, SLM -- spatial light modulator.}
\end{figure*}

\begin{figure}[!t]
	\includegraphics[width=0.49\textwidth]{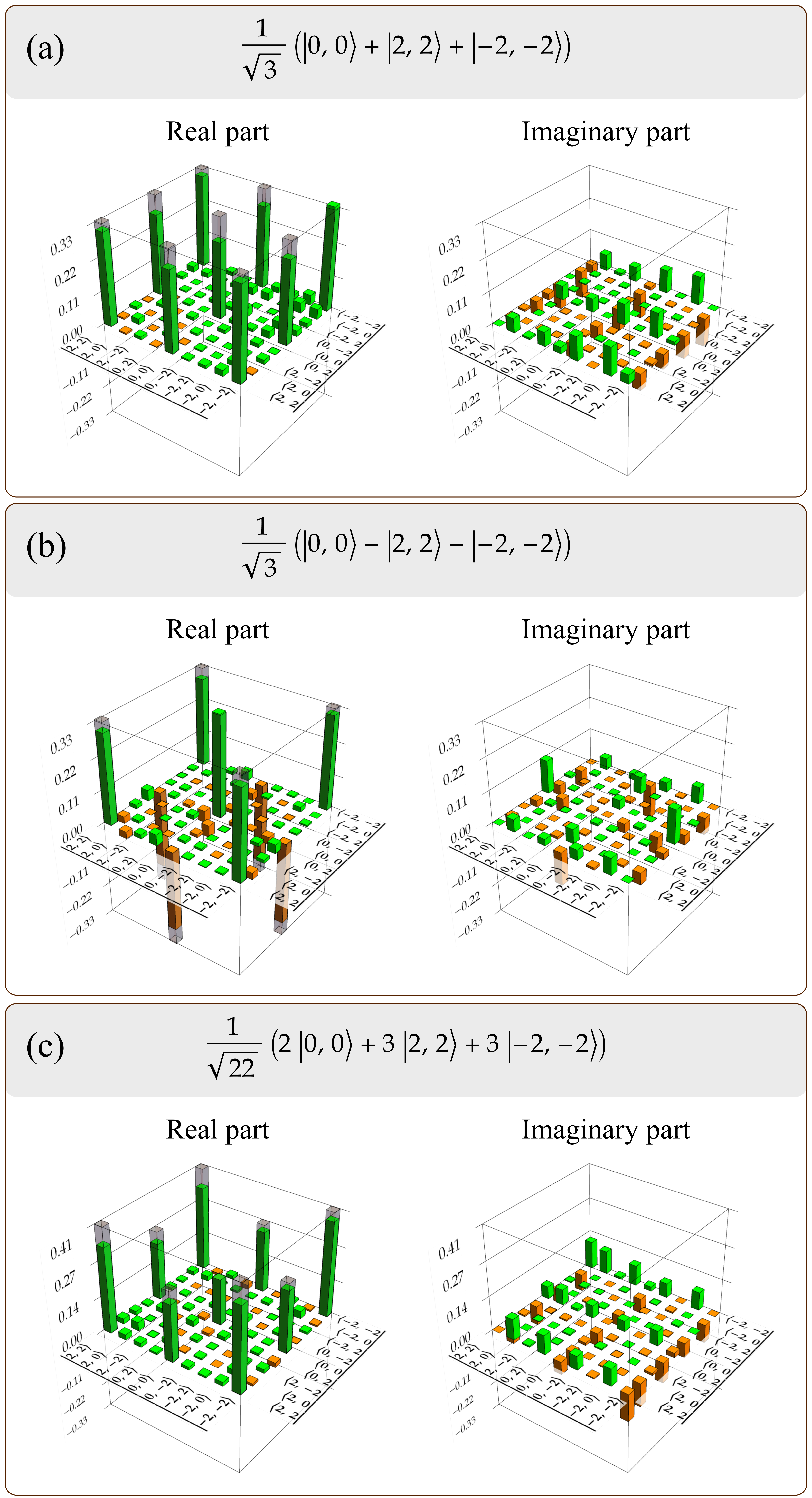}
	\caption{Three selected three-dimensionally entangled states. Real and imaginary parts of density matrices are shown. Green and orange bars represent positive and negative values of reconstructed density matrices, respectively. Gray bars represent the theoretical expectation. Fidelities of the measured states with their reference states are $87.0 \pm 0.5 \%$, $89.0 \pm 0.4 \%$ and $84.8 \pm 0.8 \%$, respectively. (a) State $\ket{\psi_1} = 1/\sqrt{3}(\ket{0,0}+\ket{2,2}+\ket{-2,-2})$. (b) State $\ket{\psi_4} = 1/\sqrt{3}(\ket{0,0}-\ket{2,2}-\ket{-2,-2})$. (c) State $\ket{\psi_5} = 1/\sqrt{22}(2 \ket{0,0} + 3 \ket{2,2} + 3 \ket{-2,-2})$.}
	\label{fig:3d_density_matrices_ReIm}
\end{figure}

\begin{figure}[htbp]
	\includegraphics[width=0.49\textwidth]{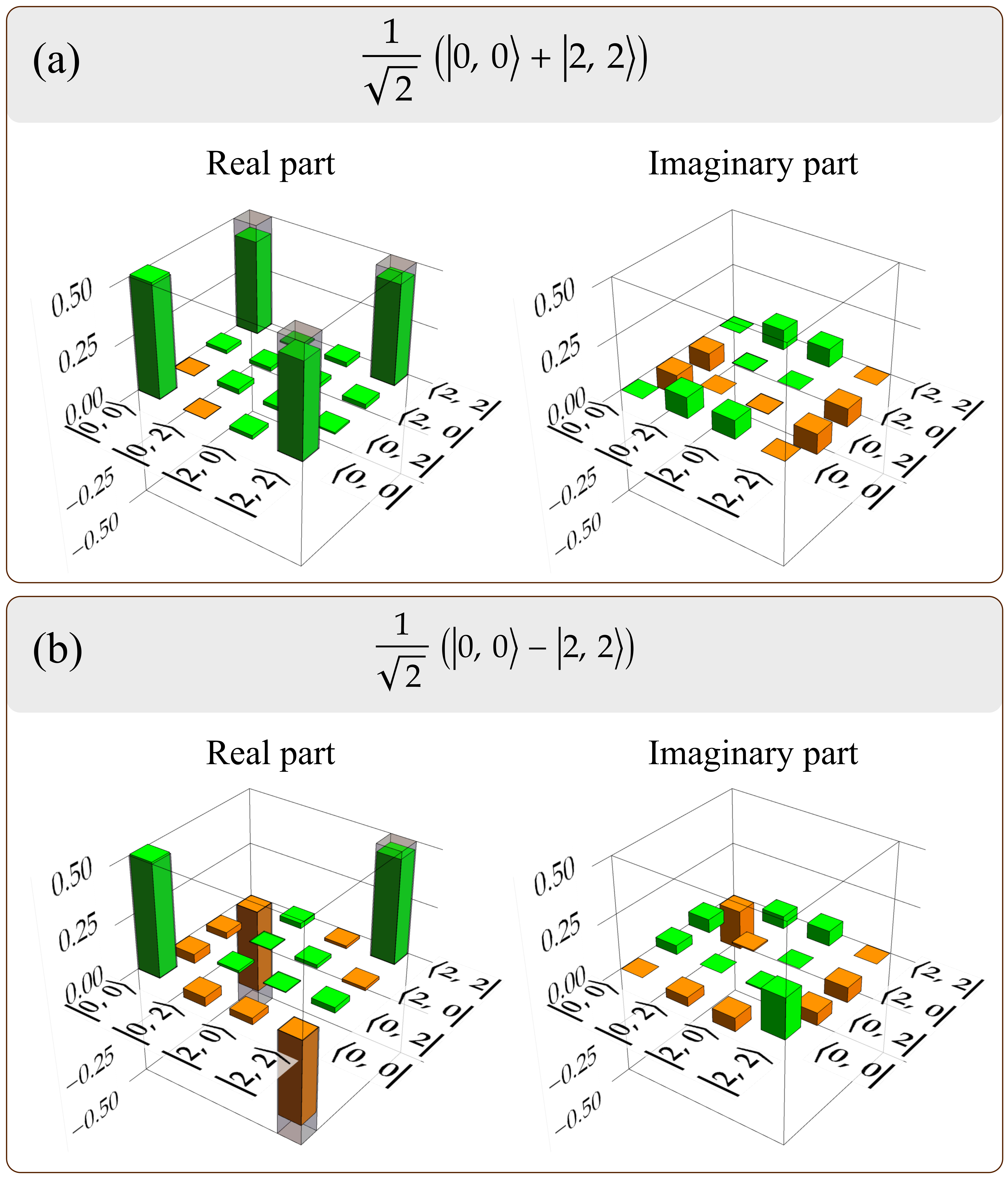}
	\caption{Two selected two-dimensionally entangled states. Real and imaginary parts are shown. Green and orange bars represent positive and negative values of reconstructed density matrices, respectively. Gray bars represent the theoretical expectation. (a) State $\ket{\Phi^+} = 1/\sqrt{2}(\ket{0,0}+\ket{2,2})$ with fidelity $90.4 \pm 0.5 \%$. (b) State $\ket{\Phi^-} = 1/\sqrt{2}(\ket{0,0}-\ket{2,2})$ with fidelity $89.1 \pm 0.5 \%$.}
	\label{fig:2d_density_matrices_reim}
\end{figure}

\begin{figure}
\includegraphics[scale=.5]{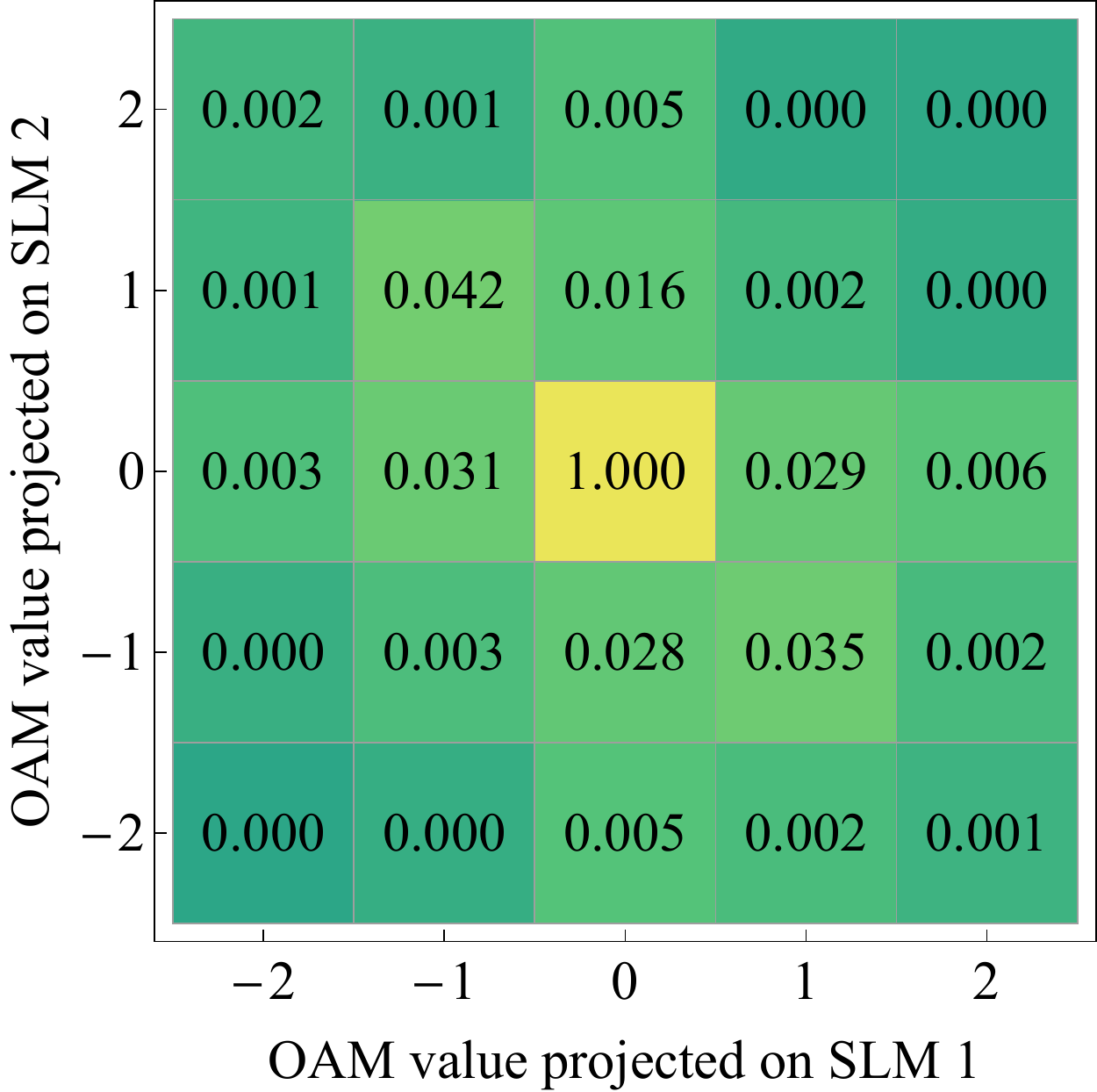}
\caption{\label{fig:spiralspectrum}%
Normalized coincidences for different projective measurements of OAM in the computational basis. Photon pairs produced in crystal A, pumped with a beam having zero quanta of OAM, were detected and coincidence counts collected for different choices of projections. On the two SLMs, see the detection system in Fig.~\ref{fig:detailed_setup}, the wavefronts corresponding to OAM modes $-2$, $\ldots$, $2$ were projected. In the ideal case, only diagonal entries would be nonzero. The coincidence rate for $\ket{0,0}$ mode is more than twenty times higher than the next highest coincidence rate.
}
\end{figure}

\section{Coherence conditions}

The generation of quantum states via the concept of entanglement by path identity requires coherent and indistinguishable photon-creation processes. To verify a sufficient level of coherence in our setup, the spiral phase plate was removed from the setup in Fig.~\ref{fig:detailed_setup} and the interference between different SPDC processes in the zero OAM mode was measured. The quality of the coherence is quantified by the interferometric visibility $V = (\text{Max}(D) - \text{Min}(D))/(\text{Max}(D) + \text{Min}(D))$, where $D$ is the coincidence count rate. Results for crystals A and B are shown in Fig.~\ref{fig:interference}. The observed visibility exceeds 97 \% in this case and the two SPDC processes in crystals A and B thus exhibit a high degree of coherence. Analogous results were also obtained for crystals B and C.

In general, the following relation has to be satisfied in order to observe interference for collinear SPDC processes. Let $L_{\mathrm{coh}}$ be the coherence length of the pump laser, which is in our case greater than 2 cm. Moreover, let $L_{\mathrm{p, A}}$ and $L_{\mathrm{p, B}}$ be the distances traveled by the pump beam from the beam splitter to crystals A and B, respectively. The physical conditions for coherence of corresponding SPDC processes are then given by \cite{coherencecond1,coherencecond2}
\begin{equation}
\abs{L_{\mathrm{p, B}} - L_{\mathrm{p, A}} - L_{\mathrm{SPDC}}} \leq L_{\mathrm{coh}},
\end{equation}
where $L_{\mathrm{SPDC}}$ is the propagation distance of down-converted photons from crystal A to crystal B. In other words, the optical path length difference between the two arms of the interferometer must be within the coherence length of the pump laser. 
\section{State tomography results}

In Fig.~\ref{fig:3d_density_matrices_ReIm} the real and imaginary parts of density matrices are shown that correspond to states presented in Fig.~3 in the main text. Analogously, in Fig.~\ref{fig:2d_density_matrices_reim} the real and imaginary parts of two-dimensional states presented in Tab.~I in the main text are shown.

\section{Spiral spectrum}

Typically, the state of photon pairs $\ket{\psi}$ produced in an SPDC process contains a non-negligible admixture of higher-order OAM terms
\begin{multline}
\ket{\psi} = \alpha_0 \ket{0,0} + \alpha_1 (\ket{1,-1} + \ket{-1,1}) + \\ \alpha_2 (\ket{2,-2} + \ket{-2,2}) + \ldots
\end{multline}
Magnitudes of these contributions in general decrease for increasing OAM order. The precise relationship between the OAM order $k$ and its complex amplitude $\alpha_k$ is governed by several tunable parameters \cite{miatto}. In order for our scheme, presented in the main text, to work properly, these parameters have to be chosen such that all higher-order OAM terms coming from the SPDC processes are significantly suppressed. As shown in Fig.~\ref{fig:spiralspectrum} for crystal A, we were able to suppress the probability $|\alpha_1|^2$ of detecting the photons in the first OAM order below five percent of the probability of detecting them in the zero mode $|\alpha_0|^2$. Similar results were obtained for crystals B and C as well. These data justify our assumption in the main text that SPDC process produces photons only in their zero OAM mode.

\section{Phase adjustment}

Relative phases in generated quantum states can be tuned precisely by a series of \mbox{quarter-,} half-, and quarter-wave plates, henceforth referred to as the QHQ scheme, that are inserted into the locking-laser beam as shown in Fig.~\ref{fig:detailed_setup}. The QHQ scheme manipulates the local phase between the horizontal and vertical polarization components of the locking-laser beam. At the polarizing beam-splitter the relative phase between polarizations translates into relative phase between the two modes of propagation of the locking-laser beam through the interferometer. After recombination of the two paths at the dichroic mirror the intensity of the interfering beam is measured by a photodiode, which feeds the measured signal to the PID controller. The controller interprets the intensity change as unwanted fluctuation and offsets the piezo actuator to compensate for it. This way the phase change is imprinted into the pump beam and therefore into the down-converted photons as well. 

When quarter-wave plates in the QHQ scheme are rotated correctly, the middle half-wave plate alone can be turned to adjust conveniently the phase in generated quantum states. In what follows, the working principle of QHQ scheme is explained. 

In Jones matrix formalism, a quarter-wave plate (Q) and a half-wave plate (H), rotated by angle $\alpha$ with respect to the vertical direction, are represented by
\begin{eqnarray}
Q(\alpha) & = & R(\alpha) \, \begin{pmatrix}
1 & 0 \\
0 & i
\end{pmatrix} \, R(-\alpha), \\
H(\alpha) & = & R(\alpha) \ \sigma_Z\ R(-\alpha)
\end{eqnarray}
respectively, where $R(\alpha)$ is a rotation matrix and $\sigma_Z$ is Pauli-Z matrix. Their forms read
\begin{equation}
R(\alpha) = \begin{pmatrix}
\cos(\alpha) & -\sin(\alpha) \\
\sin(\alpha) & \cos(\alpha)
\end{pmatrix}, \
\sigma_Z = \begin{pmatrix}
1 & 0 \\
0 & -1
\end{pmatrix}.
\end{equation}

We first explain the working principle of the QHHQ scheme, where two half-wave plates are used, and then show that this scheme is equivalent to the QHQ scheme. It can be shown that a single Q and a single H can be used to transform any elliptical polarization into a linear polarization. Such a linear polarization can then be easily rotated by a half-wave plate independently of the input polarization. Finally, a quarter-wave plate rotated by $\pi/4$ transforms a linearly polarized state with polarization angle $\varphi$ into an equally-weighted superposition of horizontal and vertical polarization components as
\begin{equation}
Q\left(\frac{\pi}{4}\right) \begin{pmatrix}
\cos(\varphi) \\ \sin(\varphi)
\end{pmatrix} = \frac{e^{i(\frac{\pi}{4}-\varphi)}}{\sqrt{2}} \begin{pmatrix}
1 \\ e^{i (2 \varphi - \frac{\pi}{2})}
\end{pmatrix}.
\label{eq:qwaveplate}
\end{equation}
The polarization angle $\varphi$ is therefore transformed into a relative phase. In total, $Q(\frac{\pi}{4}) H(\alpha) H(\beta) Q(\gamma)$ scheme allows one to obtain a beam with polarization of the form $H + e^{i \omega} V$, where $\beta$ and $\gamma$ depend on the input polarization as generated by the locking laser and relative phase $\omega$ depends effectively only on the rotation angle $\alpha$ of the half-wave plate.\\[1ex]

It is straightforward to prove two useful relations $H(\alpha) H(\beta) = H(\alpha - \beta) \sigma_Z$ and $\sigma_Z Q(-\gamma) = Q(\gamma) \sigma_Z$, so that 
\begin{equation}
Q \left( \frac{\pi}{4} \right) H(\alpha) H(\beta) Q(\gamma) = Q \left(\frac{\pi}{4} \right) H(\alpha - \beta) Q(-\gamma) \sigma_Z.
\label{eq:qhq}
\end{equation}
The extra $\sigma_Z$ merely shifts the relative phase of the incoming beam by $\pi$, which is corrected for by the proper setting of $\beta$ and $\gamma$. We thus showed that $Q(\frac{\pi}{4}) H(\alpha - \beta) Q(-\gamma)$ scheme can be used to adjust the relative phase in the state of the locking-laser beam by turning the half-wave plate appropriately.

\bibliography{references}